# V0784 Ophiuchi: an RR Lyrae star with multiple Blazhko modulations

Pierre de Ponthière [1]
*15 Rue Pré Mathy, Lesve – Profondeville 5170 - Belgium*

Franz-Josef (Josch) Hambsch [1,2,3]
*12 Oude Bleken, Mol, 2400 - Belgium*

Tom Krajci [1]
*P.O. Box 1351, Cloudcroft, NM 88317 - USA*

Kenneth Menzies [1]
*318A Potter Road, Framingham MA,01701 - USA*

1 American Association of Variable Star Observers (AAVSO)
2 Bundesdeutsche Arbeitsgemeinschaft für Veränderliche Sterne e.V. (BAV), Germany
3 Vereniging Voor Sterrenkunde (VVS), Belgium

## Abstract

The results of an observation campaign of V0784 Ophiuchi over a time span of two years have revealed a multi-periodic Blazhko effect. A Blazhko effect for V0784 Ophiuchi has not been reported previously. From the observed light curves, 60 pulsation maxima have been measured. The Fourier analyses of the (O-C) values and of magnitudes at maximum light ($M_{max}$) have revealed a main Blazhko period of 24.51 days but also two other secondary Blazhko modulations with periods of 34.29 and 31.07 days. A complete light curve Fourier analysis with PERIOD04 has shown triplet structures based on main and secondary Blazhko frequencies close to the reciprocal of Blazhko periods measured from the 60 pulsation maxima.

## 1. Introduction

In the General Catalogue of Variable Stars (Samus *et al.* 2011) V0784 Oph is correctly classified as an RR Lyr (RRab) variable star but with an incorrect period of 0.3762746 day as pointed out by Wils *et al.* (2006). They corrected the period to 0.60336 days but did not detect a Blazhko effect from the Northern Sky Variability Survey data (Wozniak *et al.* 2004) due to the paucity of available datasets.

The current data were gathered during 238 nights between June 2011 and October 2012. During this period of 488 days, a total of 15,223 magnitude measurements covering 18 Blazhko cycles were collected. The observations were made by Krajci and Hambsch using 30 cm and 40 cm telescopes located in Cloudcroft (New Mexico), Mol (Belgium), and mainly in San Pedro de Atacama (Chile). The numbers of observations for the different locations are 1,986 for Cloudcroft, 470 for Mol and 12,767 for San Pedro de Atacama.

The comparison stars are given in Table 1. The star coordinates and magnitudes in B and V bands were obtained from the NOMAD catalogue (Zacharias *et al.* 2011). C1 was used as a magnitude reference and C2 as a check star. The Johnson V magnitudes from different instruments have not been transformed to the standard system since measurements were performed with a V filter only. However, two simultaneous maximum measurements from the instruments in Cloudcroft and San Pedro de Atacama were observed to differ by only 0.034 and 0.042 mag. Dark and flat field corrections were performed with MAXIMDL software (Diffraction Limited, 2004), and aperture photometry was performed using LESVEPHOTOMETRY (de Ponthière, 2010), a custom software which also evaluates the SNR and estimates magnitude errors.

## 2. Light curve maxima analysis

The times of maxima of the light curves have been evaluated with custom software (de Ponthiere, 2010) fitting the light curve with a smoothing spline function (Reinsch, 1967). Table 2 provides the list of 60 observed maxima and Figures 1a and 1b show the (O-C) and $M_{max}$ (Magnitude at Maximum) values. For clarity only the most intensive part of the observation campaign is included in Figures 1a and 1b. From a simple inspection of the $M_{max}$ graph the Blazhko effect is obvious as well as the presence of a second modulation frequency. The Blazhko effect is itself apparently modulated by a lower frequency component.

A linear regression of all available (O-C) values has provided a pulsation period of 0.6033557 d (1.657397 $d^{-1}$). The (O-C) values have been re-evaluated with this new pulsation period and the pulsation ephemeris origin has been set to the highest recorded maximum: HJD 456 047.7942. The new derived pulsation elements are:

$$HJD_{Pulsation} = (2\,456\,047.7942 \pm 0.0015) + (0.6033557 \pm 0.0000078)\,E_{Pulsation}$$

The derived pulsation period is in good agreement with the value of 0.60336 published by Wils *et al*. (2006). The folded light curve on this pulsation period is shown in Figure 2.

To determine the Blazhko effect, Fourier analyses and sine-wave fittings of the (O-C) values and $M_{max}$ (Magnitude at Maximum) values were performed with PERIOD04 (Lenz and Breger 2005). These analyses were limited to the first two frequency components and are tabulated below:

Blazhko spectral components from (O-C) values

| Frequency (cycle/days) | $\sigma(d^{-1})$ | Period (days) | $\sigma(d)$ | Amplitude (days) | Φ (cycle) |
|---|---|---|---|---|---|
| 0.04071 | 13 E-5 | 24.56 | 0.08 | 0.0098 | 0.502 |
| 0.02916 | 14 E-5 | 34.29 | 0.17 | 0.0091 | 0.195 |

Blazhko spectral components from $M_{max}$

| Frequency (cycle/days) | $\sigma(d^{-1})$ | Period (days) | $\sigma(d)$ | Amplitude (mag) | Φ (cycle) |
|---|---|---|---|---|---|
| 0.04080 | 4 E-5 | 24.51 | 0.02 | 0.229 | 0.622 |
| 0.03219 | 8 E-5 | 31.07 | 0.08 | 0.112 | 0.620 |

The frequency uncertainties have been evaluated from the Least-Square fittings. The obtained periods (24.56 and 24.51 days) for the first Blazhko effect agree within the errors. However, the periods (34.29 and 31.07 days) of the second Blazhko modulation are statistically different and in the next section it will be shown that they are also found in the complete light curve Fourier analysis.
To dismiss the possible effect of minor amplitude differences between the non-standardized observations, the Blazhko spectral components from $M_{max}$ have been re-evaluated with only the observations from San Pedro de Atacama (Chile). The resulting frequencies (0.04081 and 0.03215 c/d) are within the uncertainties of the complete dataset.

The (O-C) and magnitude at maximum curves folded with the Blazhko period are given in Figure 3a and 4a. In these diagrams, the scatter of the data is mainly due to the presence of the second Blazhko modulation. These data were folded using the period of 24.56 days for the (O-C) values and 24.51 days for the $M_{max}$ values. When a pre-whitening with the frequency of 0.02916 c/d is applied to the (O-C) dataset, the scatter is reduced significantly as is shown in Figure 3b. The scatter of the $M_{max}$ data is also reduced after a pre-whitening with the corresponding frequency of 0.03219 c/d (Figure 4b).

On this basis the best Blazhko ephemeris is

$$HJD_{Blazhko} = 2456047.7942 + (24.51 \pm 0.02) \, E_{Blazhko}$$

where the origin has been selected as the epoch of the highest recorded maximum. The peak-to-peak magnitude variation during the night of highest recorded maximum is 1.34 mag. Over the Blazhko period, the magnitude at maximum brightness differs by about 0.66 mag, that is, 49% of the light curve peak-to-peak variations. The (O-C) values differ in a range of 0.046 day, that is, 7.7% of the pulsation period.

## 3. Frequency spectrum analysis of the light curve

From the light curve maxima analysis, the pulsation and Blazhko frequencies and other frequencies modulating the Blazhko effect have been identified. It is interesting to note that these modulating frequencies are clearly present in the spectrum of the complete light curve.

The Blazhko effect can be seen as an amplitude and phase modulation of the periodic pulsation with the modulation frequency being the Blazhko frequency (Szeidl and Jurcisk 2009). The spectrum of a signal modulated in amplitude and phase is characterized by a pattern of peaks called multiplets at the positions $kf_0 \pm nf_B$ with k and n being integers corresponding respectively to the harmonic and multiplet orders. The frequencies, amplitudes and phases of the multiplets can be measured by a succession of Fourier analyses, pre-whitenings and sine-wave fittings.
This analysis has been performed with PERIOD04. Only the harmonic and multiplet components having a signal to noise ratio (SNR) greater than 4 have been retained as significant signals.

Besides the pulsation frequency $f_0$, harmonics $nf_0$ and series of triplets $nf_0 \pm f_B$ based on the principal Blazhko frequency $f_B$, other triplets have also been found and are tabulated below :

Triplet component frequencies and periods

| Component | Derived from | Frequency ($d^{-1}$) | $\sigma(d^{-1})$ | Period (d) | $\sigma(d)$ |
|---|---|---|---|---|---|
| $f_0$ |  | 1.657399 | 3.2 E-6 | 0.6033551 | 1.16 E-6 |
| $f_B$ | $f_0 + f_B$ | 0.040544 | 29 E-6 | 24.66 | 0.02 |
| $f_{B2}$ | $f_0 + f_{B2}$ | 0.028820 | 35 E-6 | 34.70 | 0.04 |
| $f_{B3}$ | $f_0 - f_{B3}$ | 0.033346 | 124 E-6 | 29.99 | 0.11 |

They are based on two other secondary modulation frequencies $f_{B2}$ and $f_{B3}$ which are close to the secondary modulation components identified in the (O-C) and $M_{Max}$ analyses.

It is interesting to note that the period (1 / $f_{B2}$) of 34.70 days is close to the second period 34.29 days found in the (O-C) analysis. The period (1 / $f_{B3}$) of 29.99 days can be compared to the second period 31.07 days found in the magnitude at maximum ($M_{Max}$) analysis. Table 3 provides the complete list of Fourier components with their amplitudes, phases and uncertainties. During the sine-wave fitting, the fundamental frequency $f_0$ and largest triplets $f_0 + f_B$, $f_0 + f_{B2}$ and $f_0 - f_{B3}$ have been left unconstrained and the other frequencies have been entered as combinations of these four frequencies. The uncertainties of frequencies, amplitudes and phases have been estimated by Monte Carlo simulations. The amplitude and phase uncertainties have been multiplied by a factor of two as it is known that the Monte Carlo simulations underestimate these uncertainties (Kolenberg *et al*. 2009). The two Blazhko modulation frequencies $f_B$ (0.040544) and $f_{B2}$ (0.028820) are close to a 7:5 resonance ratio and the corresponding beat period is around 173 days. Calculations based on Blazhko periods (24.56 and 34.29 days) obtained with the (O-C) analysis provide the same resonance ratio. The other pair of frequencies $f_B$ (0.040544) and $f_{B3}$ (0.033346) is close to a 5:6 resonance ratio with a beat period around 149 days, but a different resonance ratio of 5:4 with a beat period of 123 days is obtained from the periods (24.51 and 31.07 days) found in the magnitude at maximum analysis. This discrepancy is probably due to the greater uncertainty (124 E-6) of the $f_{B3}$ side lobes. In CZ Lacertae, Sódor *et al*.(2011) have also detected two modulation components with a resonance ratio of 5:4 during a first observation season but a different resonance ratio of 4:3 during the next season.

Table 4 lists for each harmonic the amplitude ratios $A_i/A_1$ and the ratios usually used to characterize the Blazhko effect, that is, $A_i^+/A_1$ ; $A_i^-/A_1$ ; $R_i = A^+_{i1} / A^-_{i1}$ and asymmetries $Q_i = (A^+_{i1} - A^-_{i1}) / (A^+_{i1} + A^-_{i1})$. Szeidl and Jurcsik (2009) have shown that the asymmetry of the side lobe amplitudes depends on the phase difference between the amplitude and the phase modulation. If the Blazhko effect is limited to amplitude or phase modulation the asymmetry vanishes. In the present case the side lobe $A^+_{i1}$ is the strongest one, which is generally the case for stars showing a Blazhko effect. The asymmetry ratios $Q_i$ around 0.32 are a sign of both strong amplitude and phase modulations. The $R_i$ and $Q_i$ ratios for triplets around the secondary Blazhko frequency $f_{B2}$ are also given in Table 4. The $f_{B2}$ frequency is close to the secondary frequency detected in the (O-C) analysis and thus likely related to a phase modulation effect. The $f_{B2}$ frequency seems to act only on phase modulation and this could explain the low asymmetry values (0.03) of the $f_{B2}$ side lobes.

## 4. Light curve variations over Blazhko cycle.

Subdividing the data set into temporal subsets is a classical method to analyze the light curve variations over the Blazhko cycle. Ten temporal subsets corresponding to the different Blazhko phase intervals $\Psi_i$ (i = 0 , 9) have been created using the epoch of the highest recorded maximum (2456047.7942) as the origin of the first subset. Fortunately, the data points are relatively well distributed over the subsets with the number of data points varying between 1,126 and 1,715. Figure 5 presents the folded light curve for the ten subsets. Despite the subdivision over the Blazhko cycle, a scatter still remains on the light curves. The light curves are folded with the primary Blazhko frequency and the secondary Blazhko frequencies are creating the observed scatter.

Fourier analyses and Least-Square fittings have been performed on the different temporal subsets. For the fundamental and the first four harmonics the amplitude $A_i$ and the epoch-independent phase differences ($\Phi_{k1} = \Phi_k - k\Phi_{k1}$) are given in Table 5 and plotted in Figure 6. The number of data points belonging to each subset is also given in this table. The amplitudes of the fundamental and harmonics show smooth sinusoidal variations with the minima occurring around Blazhko phase 0.5, that is, when the light curve amplitude variation on the pulsation is weaker. The ratio of harmonic $A_4$ to fundamental $A_1$ is at maximum at Blazhko phase 1.0, when the ascending branch of the light curve is steeper. The difference between maximum and minimum $\Phi_1$ phases is a measure of the phase modulation strength and is equal to 0.453 radian or 0.072 cycle, which corresponds roughly to the value of 7.7% noted for the peak to peak deviation of (O-C). The epoch-independent phase differences $\Phi_{31}$ and $\Phi_{41}$ vary as smooth sinusoids, with the maximum phase differences occurring at Blazhko phase 0.5. However $\Phi_{21}$ variations, if any, are small. This weak $\Phi_{21}$ variation has been also noted for MW Lyr (Jurcsik *et al*. 2008) and V1820 Ori (de Ponthière *et al*. 2013).

## 5. Conclusions

The effects of three Blazhko modulations have been detected by measurements of (O-C) values and amplitude of light curve maxima and confirmed by complete light curve Fourier analysis. The main Blazhko period (1 / $f_B$) is 24.51 days. The secondary Blazhko period (1 / $f_{B2}$) of 34.70 days is apparently related to phase modulation as it is detected in the (O-C) analysis, and $f_B$ and $f_{B2}$ are close to a 7:5 resonance ratio. The tertiary modulation (1 / $f_{B3}$) is weaker in the Fourier analysis and the period values are slightly different from magnitude at maximum and light curve Fourier analysis (31.07 and 29.99 days). The resonance ratios of $f_B$ and $f_{B3}$ are approximately 5:4 or 5:6 in function of the analysis method. This discrepancy is probably due to the weakness of the corresponding Fourier multiplet values and their larger uncertainties.

## Acknowledgements

Dr A. Henden, Director of AAVSO and the AAVSO are acknowledged for the use of AAVSOnet telescopes at Cloudcroft (New Mexico, USA). The authors thank the referee for constructive comments which have helped to clarify and improve the paper. This work has made use of The International Variable Star Index (VSX) maintained by the AAVSO and of the SIMBAD astronomical database (http://simbad.u-strasbg.fr)## References

de Ponthière, P. 2010, LESVEPHOTOMETRY, automatic photometry software (http://www.dppobservatory.net).
de Ponthière, P. *et al*. 2013, *J. Amer. Assoc. Var. Star Obs*. accepted for publication. (http://arxiv.org/abs/1212.0897)
Diffraction Limited. 2004, MAXIMDL image processing software (http://www.cyanogen.com).
Jurcsik, J., *et al*. 2008, *Mon. Not. Roy. Astron. Soc.*, **391**, 164.
Kolenberg, K., *et al*. 2009, *Mon. Not. Roy. Astron. Soc.*, **396**, 263.
Lenz, P., and Breger, M. 2005, *Commun. Asteroseismology*, **146**, 53.
Reinsch, C. H. 1967, *Numer. Math.*, **10**, 177.
Samus, N. N., *et al*. 2011 *General Catalogue of Variable Stars* (GCVS database, Version 2011 January, http://www.sai.msu.su/gcvs/gcvs/index.htm).
Sódor *et al*. 2011, *Mon. Not. Roy. Astron. Soc.*, **411**, 1585.
Szeidl, B., and Jurcsik, J. 2009 *Commun. Asteroseismology*, **160**, 17.
Wils, P., Lloyd, C., Bernhard, K. 2006, *Mon. Not. Roy. Astron. Soc.*, **368**, 1757.
Wozniak, P. *et al*. 2004, *Astron. J.*, **127**, 2436
Zacharias, N., Monet, D., Levine, S., Urban, S., Gaume, R., and Wycoff, G. 2011, The Naval Observatory Merged Astrometric Dataset (NOMAD, http://www.usno.navy.mil/USNO/astrometry/optical-IR-prod/nomad/).

**Table 1. Comparison stars for V0784 Ophiuchi**

| Identification | R.A. (2000) h m s | Dec (2000) ° ' " | B | V | B-V | |
|---|---|---|---|---|---|---|
| GSC 992-1617 | 17 35 18.037 | +07 40 58.97 | 13.24 | 12.32 | 0.92 | C1 |
| TYC 992-1203 | 17 35 35.883 | +07 41 29.21 | 11.491 | 10.442 | 1.049 | C2 |

**Table 2. List of measured maxima of V0784 Ophiuchi**

| Maximum HJD | Error | O-C (day) | E | Magnitude | Error | Location* |
|---|---|---|---|---|---|---|
| 2455778.6910 | 0.0014 | -0.0066 | -446 | 11.692 | 0.004 | 1 |
| 2455780.5024 | 0.0021 | -0.0052 | -443 | 11.841 | 0.004 | 1 |
| 2455783.5146 | 0.0035 | -0.0098 | -438 | 12.043 | 0.005 | 1 |
| 2455792.5913 | 0.0055 | 0.0166 | -423 | 11.952 | 0.022 | 1 |
| 2455795.6064 | 0.0023 | 0.0149 | -418 | 11.757 | 0.005 | 1 |
| 2455798.6121 | 0.0022 | 0.0038 | -413 | 11.581 | 0.004 | 1 |
| 2455801.6283 | 0.0012 | 0.0032 | -408 | 11.580 | 0.004 | 1 |
| 2455804.6413 | 0.0018 | -0.0006 | -403 | 11.747 | 0.004 | 1 |
| 2455815.5203 | 0.0062 | 0.0180 | -385 | 12.139 | 0.005 | 1 |
| 2455818.5302 | 0.0040 | 0.0112 | -380 | 11.998 | 0.005 | 1 |
| 2455989.8824 | 0.0030 | 0.0103 | -96 | 11.897 | 0.006 | 1 |
| 2455992.8923 | 0.0021 | 0.0035 | -91 | 11.762 | 0.007 | 1 |
| 2456009.8043 | 0.0036 | 0.0215 | -63 | 11.971 | 0.008 | 1 |
| 2456012.8109 | 0.0027 | 0.01133 | -58 | 11.830 | 0.007 | 1 |
| 2456015.8195 | 0.0018 | 0.00315 | -53 | 11.683 | 0.006 | 1 |
| 2456018.8292 | 0.0015 | -0.00393 | -48 | 11.569 | 0.006 | 1 |
| 2456024.8557 | 0.0014 | -0.01098 | -38 | 11.751 | 0.009 | 1 |
| 2456027.8774 | 0.0026 | -0.00606 | -33 | 11.965 | 0.009 | 1 |
| 2456035.7338 | 0.0025 | 0.00671 | -20 | 12.028 | 0.007 | 1 |
| 2456038.7570 | 0.0033 | 0.01314 | -15 | 11.898 | 0.007 | 1 |
| 2456038.7570 | 0.0033 | 0.01314 | -15 | 11.898 | 0.007 | 1 |
| 2456047.7942 | 0.0019 | 0.00000 | 0 | 11.512 | 0.006 | 1 |
| 2456050.8100 | 0.0019 | -0.00098 | 5 | 11.753 | 0.007 | 1 |
| 2456053.8208 | 0.0025 | -0.00696 | 10 | 11.931 | 0.010 | 1 |
| 2456056.8352 | 0.0041 | -0.00934 | 15 | 12.120 | 0.008 | 1 |
| 2456059.8458 | 0.0071 | -0.01551 | 20 | 12.172 | 0.011 | 1 |
| 2456062.8785 | 0.0082 | 0.00041 | 25 | 12.091 | 0.007 | 1 |
| 2456064.6946 | 0.0034 | 0.00644 | 28 | 11.939 | 0.007 | 1 |
| 2456065.8972 | 0.0029 | 0.00233 | 30 | 11.856 | 0.008 | 1 |
| 2456067.7028 | 0.0019 | -0.00214 | 33 | 11.709 | 0.007 | 1 |
| 2456068.9082 | 0.0009 | -0.00345 | 35 | 11.624 | 0.008 | 1 |
| 2456070.7195 | 0.0016 | -0.00222 | 38 | 11.555 | 0.006 | 1 |
| 2456073.7360 | 0.0014 | -0.00249 | 43 | 11.563 | 0.007 | 1 |
| 2456076.7611 | 0.0021 | 0.00583 | 48 | 11.764 | 0.007 | 1 |
| 2456076.7625 | 0.0032 | 0.00723 | 48 | 11.806 | 0.015 | 2 |
| 2456079.7870 | 0.0026 | 0.01495 | 53 | 11.922 | 0.008 | 1 |
| 2456081.6018 | 0.0037 | 0.01968 | 56 | 11.980 | 0.010 | 1 |
| 2456082.8111 | 0.0090 | 0.02227 | 58 | 12.060 | 0.013 | 1 |
| 2456084.6218 | 0.0046 | 0.02290 | 61 | 12.036 | 0.010 | 1 |
| 2456085.8238 | 0.0052 | 0.01819 | 63 | 12.044 | 0.009 | 1 |
| 2456087.6345 | 0.0045 | 0.01882 | 66 | 11.997 | 0.008 | 1 |
| 2456088.8314 | 0.0043 | 0.00901 | 68 | 11.966 | 0.008 | 1 |
| 2456088.8321 | 0.0045 | 0.00971 | 68 | 12.000 | 0.012 | 2 |
| 2456091.8378 | 0.0024 | -0.00137 | 73 | 11.906 | 0.013 | 2 |
| 2456093.6474 | 0.0026 | -0.00183 | 76 | 11.771 | 0.007 | 1 |
| 2456096.6554 | 0.0015 | -0.01061 | 81 | 11.712 | 0.007 | 1 |
| 2456102.6891 | 0.0051 | -0.01047 | 91 | 11.870 | 0.012 | 2 |
| 2456105.7166 | 0.0036 | 0.00025 | 96 | 11.961 | 0.010 | 2 |
| 2456108.7536 | 0.0033 | 0.02048 | 101 | 11.932 | 0.009 | 1 |
| 2456110.5743 | 0.0040 | 0.03111 | 104 | 11.895 | 0.010 | 1 |
| 2456111.7694 | 0.0030 | 0.01950 | 106 | 11.824 | 0.012 | 1 |
| 2456113.5785 | 0.0027 | 0.01853 | 109 | 11.783 | 0.010 | 1 |
| 2456114.7832 | 0.0024 | 0.01652 | 111 | 11.727 | 0.009 | 1 |
| 2456116.5854 | 0.0012 | 0.00865 | 114 | 11.631 | 0.007 | 1 |
| 2456122.6012 | 0.0018 | -0.00911 | 124 | 11.673 | 0.006 | 1 |
| 2456125.6160 | 0.0028 | -0.01108 | 129 | 11.821 | 0.007 | 1 |
| 2456131.6492 | 0.0068 | -0.01144 | 139 | 12.059 | 0.008 | 1 |
| 2456134.6806 | 0.0037 | 0.00318 | 144 | 11.988 | 0.008 | 1 |
| 2456145.5391 | 0.0008 | 0.00128 | 162 | 11.572 | 0.007 | 1 |
| 2456227.5864 | 0.0027 | -0.00780 | 298 | 11.955 | 0.006 | 2 |

*Locations : 1- San Pedro de Atacama (Chile); 2 – Cloudcroft NM*

**Table 3 Multi-frequency fit results for V0784 Ophiuchi**

| Component | $f(d^{-1})$ | $\sigma(f)$ | $A_i$ (mag) | $\sigma(A_i)$ | $\Phi_i$ (cycle) | $\sigma(\Phi_i)$ | SNR |
|---|---|---|---|---|---|---|---|
| $f_0$ | 1.657399 | 3.2 E-06 | 0.3594 | 0.0017 | 0.01810 | 0.0007 | 140.8 |
| $2 f_0$ | 3.314798 | | 0.1612 | 0.0016 | 0.42518 | 0.0017 | 71.9 |
| $3 f_0$ | 4.972196 | | 0.0963 | 0.0016 | 0.87223 | 0.0024 | 47.6 |
| $4 f_0$ | 6.629595 | | 0.0587 | 0.0013 | 0.32035 | 0.0049 | 31.3 |
| $5 f_0$ | 8.286994 | | 0.0324 | 0.0017 | 0.77462 | 0.0080 | 18.1 |
| $6 f_0$ | 9.944393 | | 0.0135 | 0.0018 | 0.17084 | 0.0207 | 7.9 |
| $7 f_0$ | 11.601792 | | 0.0094 | 0.0016 | 0.56694 | 0.0289 | 6.0 |
| $8 f_0$ | 13.259190 | | 0.0060 | 0.0017 | 0.99881 | 0.0358 | 4.2 |
| $f_0 + f_B$ | 1.697943 | 29 E-06 | 0.0516 | 0.0016 | 0.62414 | 0.0047 | 20.3 |
| $f_0 - f_B$ | 1.616855 | | 0.0263 | 0.0017 | 0.40709 | 0.0113 | 10.4 |
| $2 f_0 + f_B$ | 3.355341 | | 0.0314 | 0.0017 | 0.00102 | 0.0082 | 14.0 |
| $2 f_0 - f_B$ | 3.274254 | | 0.0140 | 0.0016 | 0.77565 | 0.0204 | 6.2 |
| $3 f_0 + f_B$ | 5.012740 | | 0.0336 | 0.0018 | 0.42290 | 0.0083 | 16.6 |
| $3 f_0 - f_B$ | 4.931653 | | 0.0183 | 0.0015 | 0.21889 | 0.0130 | 9.0 |
| $4 f_0 + f_B$ | 6.670139 | | 0.0254 | 0.0016 | 0.85193 | 0.0111 | 13.6 |
| $4 f_0 - f_B$ | 6.589051 | | 0.0166 | 0.0018 | 0.67866 | 0.0166 | 8.9 |
| $5 f_0 + f_B$ | 8.327538 | | 0.0156 | 0.0015 | 0.30249 | 0.0178 | 8.7 |
| $5 f_0 - f_B$ | 8.246450 | | 0.0116 | 0.0017 | 0.11565 | 0.0214 | 6.5 |
| $6 f_0 + f_B$ | 9.984937 | | 0.0101 | 0.0016 | 0.73701 | 0.0248 | 5.9 |
| $6 f_0 - f_B$ | 9.903849 | | 0.0088 | 0.0016 | 0.60313 | 0.0297 | 5.1 |
| $7 f_0 + f_B$ | 11.642335 | | 0.0087 | 0.0016 | 0.19222 | 0.0299 | 5.5 |
| $f_0 + f_{B2}$ | 1.686218 | 35 E-06 | 0.0237 | 0.0018 | 0.77081 | 0.0114 | 9.3 |
| $f_0 - f_{B2}$ | 1.628579 | | 0.0224 | 0.0018 | 0.94771 | 0.0133 | 8.8 |
| $2 f_0 + f_{B2}$ | 3.343617 | | 0.0187 | 0.0016 | 0.19953 | 0.0119 | 8.3 |
| $2 f_0 - f_{B2}$ | 3.285978 | | 0.0175 | 0.0016 | 0.32003 | 0.0155 | 7.8 |
| $3 f_0 + f_{B2}$ | 5.001016 | | 0.0185 | 0.0017 | 0.62995 | 0.0148 | 9.1 |
| $3 f_0 - f_{B2}$ | 4.943377 | | 0.0149 | 0.0016 | 0.75635 | 0.0183 | 7.4 |
| $4 f_0 + f_{B2}$ | 6.658415 | | 0.0133 | 0.0018 | 0.08536 | 0.0206 | 7.1 |
| $4 f_0 - f_{B2}$ | 6.600776 | | 0.0152 | 0.0017 | 0.19267 | 0.0172 | 8.1 |
| $5 f_0 + f_{B2}$ | 8.315814 | | 0.0098 | 0.0017 | 0.52105 | 0.0276 | 5.5 |
| $5 f_0 - f_{B2}$ | 8.258174 | | 0.0104 | 0.0018 | 0.61650 | 0.0246 | 5.8 |
| $f_0 - f_{B3}$ | 1.624053 | 124 E-06 | 0.0178 | 0.0022 | 0.90174 | 0.0147 | 8.1 |
| $f_0 + f_{B3}$ | 1.690744 | 0.0112 | 0.0019 | 0.11617 | 0.0246 | 5.4 | |

**Table 4. V0784 Oph Harmonic, Triplet amplitudes, ratios and asymmetry parameters**

| $i$ | $A_i/A_1$ | $A_i^+/A_1$ | $A_i^-/A_1$ | $R_i$ | $Q_i$ | $R_i (f_{B2})$ | $Q_i (f_{B2})$ |
|---|---|---|---|---|---|---|---|
| 1 | 1.00 | 0.14 | 0.07 | 1.96 | 0.32 | 1.06 | 0.03 |
| 2 | 0.45 | 0.09 | 0.04 | 2.24 | 0.38 | 1.07 | 0.03 |
| 3 | 0.27 | 0.09 | 0.05 | 1.84 | 0.30 | 1.24 | 0.11 |
| 4 | 0.16 | 0.07 | 0.05 | 1.53 | 0.21 | 0.88 | -0.07 |
| 5 | 0.09 | 0.04 | 0.03 | 1.35 | 0.15 | 0.94 | -0.03 |
| 6 | 0.04 | 0.03 | 0.02 | 1.15 | 0.07 | - | - |
| 7 | 0.03 | 0.02 | - | - | - | - | - |
| 8 | 0.02 | - | - | - | - | - | - |

**Table 5. V0784 Oph Fourier coefficients over Blazhko cycles**

| $\Psi$ (cycle) | $A_1$ (mag) | $A_2$ (mag) | $A_3$ (mag) | $A_4$ (mag) | $A_4/A_1$ | $\Phi_1$ (rad) | $\Phi_{21}$ (rad) | $\Phi_{31}$ (rad) | $\Phi_{41}$ (rad) |
|---|---|---|---|---|---|---|---|---|---|
| 0.0 - 0.1 | 0.423 | 0.191 | 0.137 | 0.093 | 0.220 | 0.161 | 2.406 | 5.024 | 1.357 |
| 0.1 - 0.2 | 0.363 | 0.162 | 0.102 | 0.064 | 0.176 | 0.025 | 2.392 | 5.126 | 1.432 |
| 0.2 - 0.3 | 0.324 | 0.144 | 0.085 | 0.055 | 0.170 | 0.383 | 2.458 | 5.300 | 1.724 |
| 0.3 - 0.4 | 0.293 | 0.123 | 0.059 | 0.042 | 0.145 | 0.399 | 2.533 | 5.417 | 2.140 |
| 0.4 - 0.5 | 0.288 | 0.114 | 0.050 | 0.027 | 0.094 | 0.248 | 2.533 | 5.631 | 2.338 |
| 0.5 - 0.6 | 0.322 | 0.133 | 0.066 | 0.036 | 0.111 | 0.287 | 2.536 | 5.411 | 2.010 |
| 0.6 - 0.7 | 0.354 | 0.154 | 0.077 | 0.041 | 0.117 | 0.168 | 2.477 | 5.276 | 1.949 |
| 0.7 - 0.8 | 0.394 | 0.174 | 0.119 | 0.072 | 0.182 | 0.067 | 2.457 | 5.166 | 1.518 |
| 0.8 - 0.9 | 0.408 | 0.197 | 0.138 | 0.101 | 0.246 | 0.202 | 2.415 | 5.121 | 1.466 |
| 0.9 - 1.0 | 0.431 | 0.196 | 0.149 | 0.096 | 0.223 | -0.054 | 2.234 | 4.797 | 1.127 |

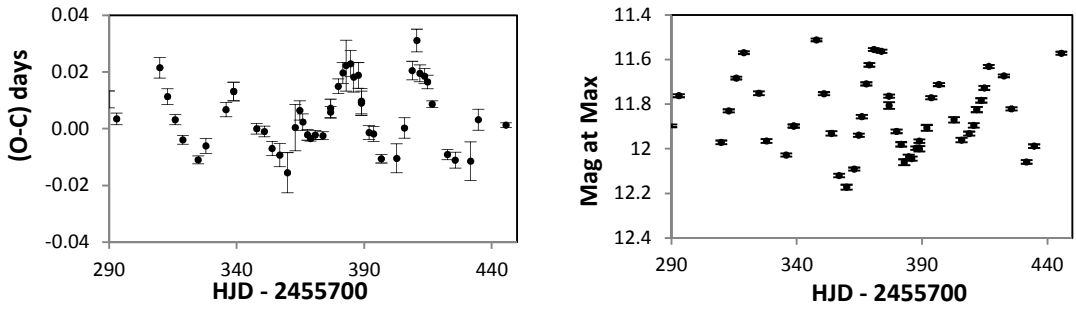

Figure 1. V0784 Oph O-C (days) and Magnitude at maximum

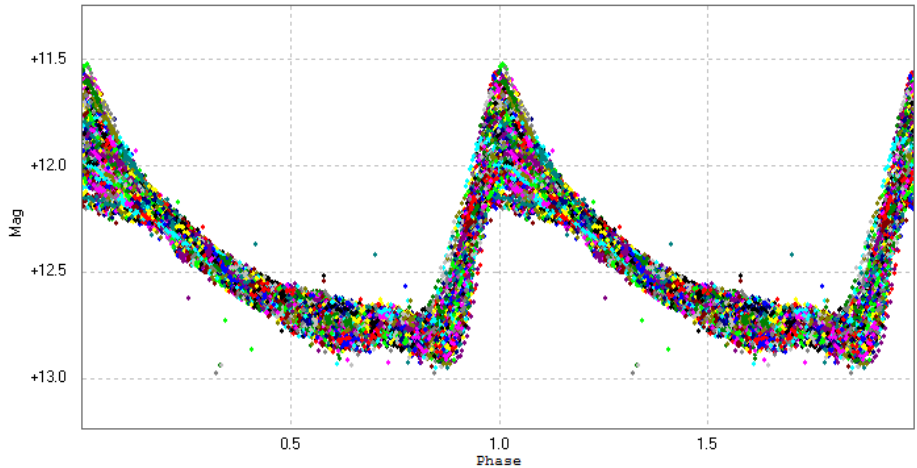

Figure 2. V0784 Oph light curve folded with pulsation period

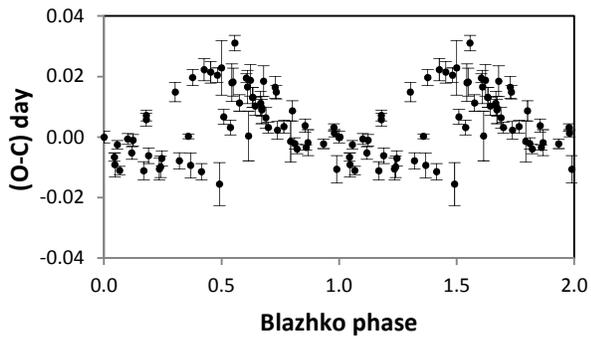
Figure 3a. O-C without pre-whitening

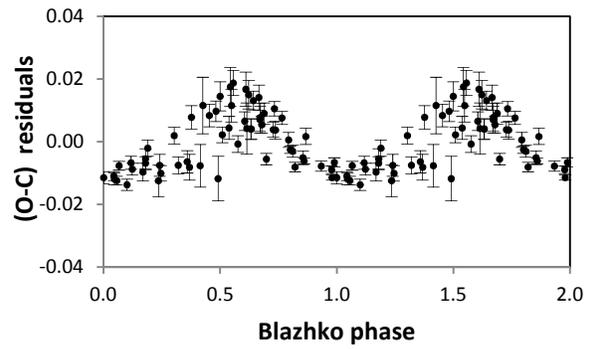
Figure 3b. O-C after pre-whitening

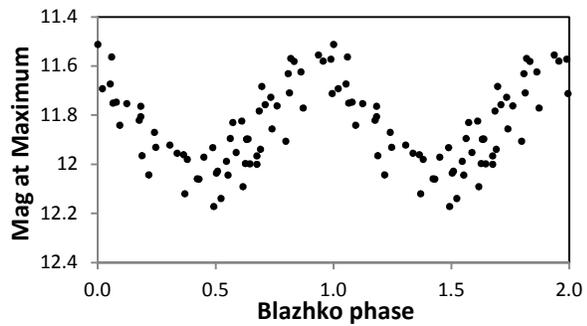
Figure 4a. Magnitude at maximum without pre-whitening

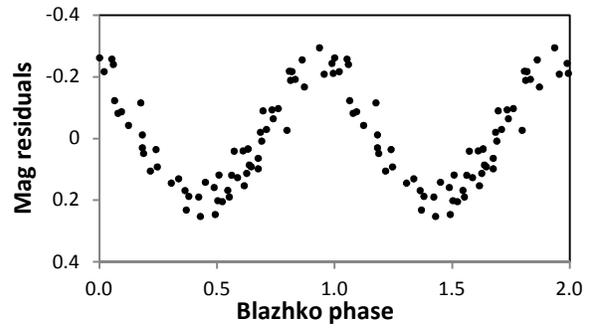
Figure 4b. Magnitude at maximum after pre-whitening

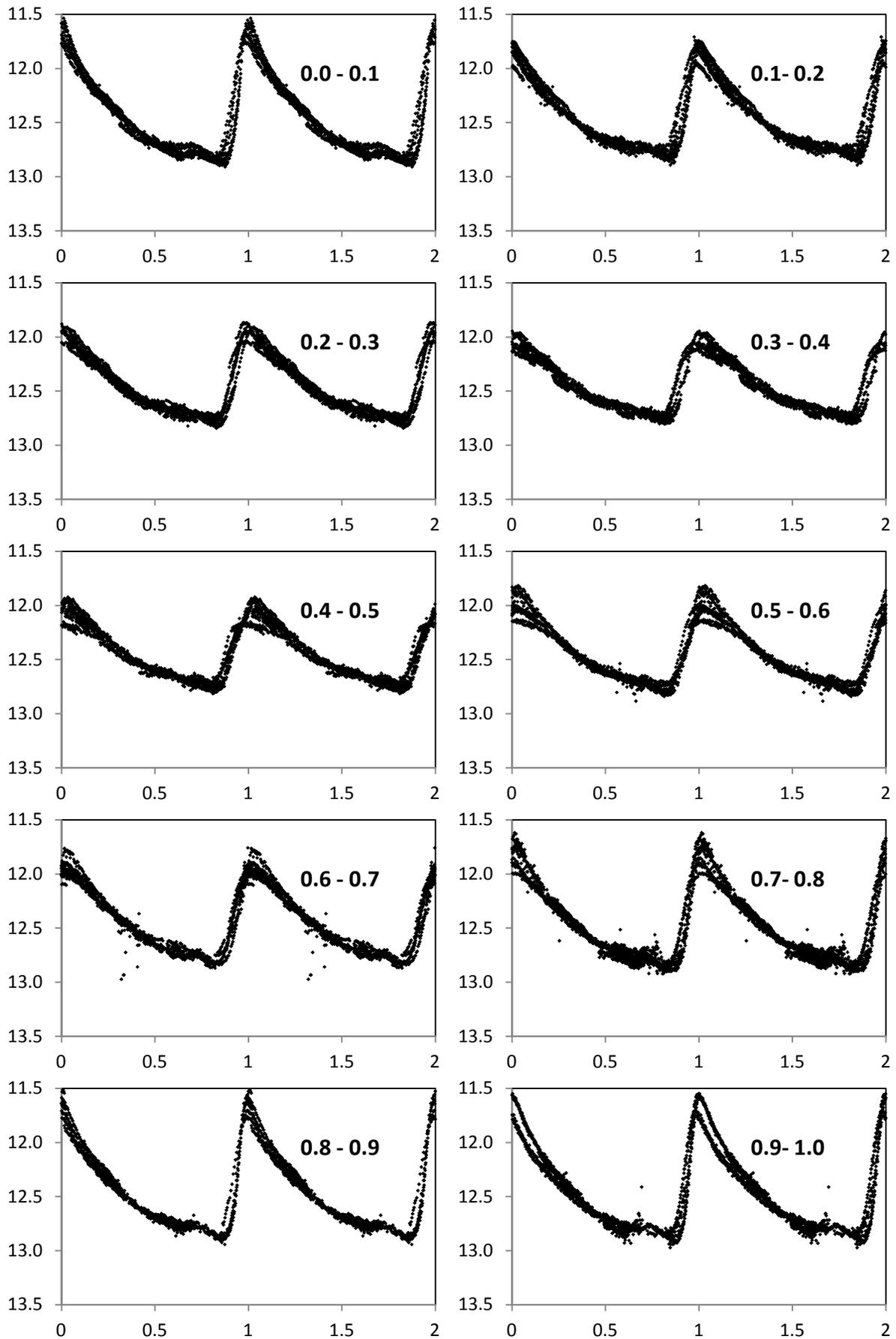
Figure 5. V0784 Oph light curve for the ten temporal subsets.

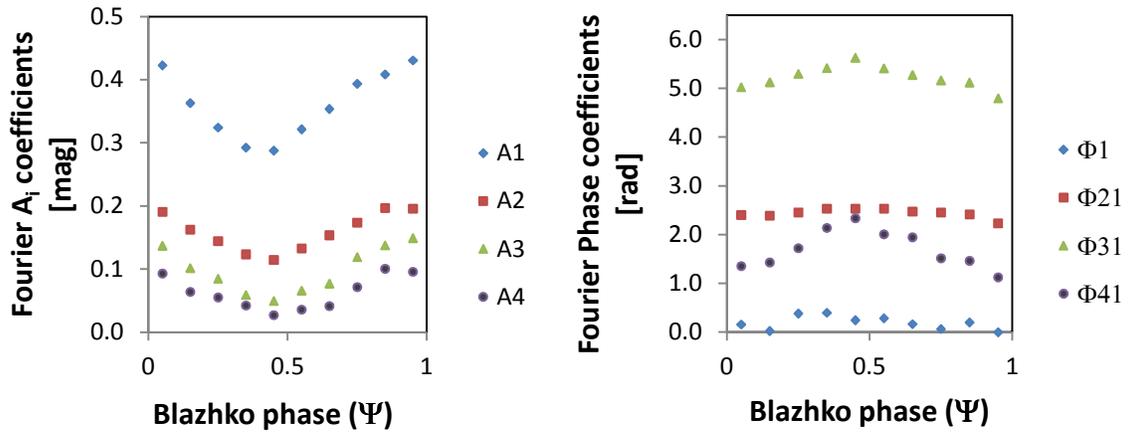

Figure 6a (left). V0784 Oph Fourier $A_i$ amplitude (mag) for the ten temporal subsets.
Figure 6b (right) Fourier $\Phi_1$ and $\Phi_{ki}$ phase (rad) for the ten temporal subsets.